\documentclass[aps,prl,twocolumn,superscriptaddress,10pt,floatfix]{revtex4-2}

\usepackage{amsmath,amssymb,bm}
\usepackage{graphicx}
\usepackage{orcidlink}
\usepackage{multirow}
\usepackage{float}

\begin{document}

\title{Space-Time Point Defects as Probes of Criticality: Application to the Gross--Neveu--Heisenberg Transition}

\author{\firstname{Jakob} \surname{V\"ath}\,\orcidlink{0009-0008-8344-2272}} 
\affiliation{Institut f\"ur Theoretische Physik und Astrophysik and W\"urzburg-Dresden Cluster of Excellence ctd.qmat, Universit\"at W\"urzburg, 97074 W\"urzburg, Germany}
\author{\firstname{Luis} \surname{Thremer}\,\orcidlink{0009-0005-9472-4548}}
\affiliation{Institut f\"ur Theoretische Physik und Astrophysik and W\"urzburg-Dresden Cluster of Excellence ctd.qmat, Universit\"at W\"urzburg, 97074 W\"urzburg, Germany}
\author{\firstname{Francesco} \surname{Parisen Toldin}\,\orcidlink{0000-0002-1884-9067}}
\affiliation{\mbox{Institute for Theoretical Solid State Physics, RWTH Aachen University, Otto-Blumenthal-Str. 26, 52074 Aachen, Germany}}
\author{\firstname{Fakher F.} \surname{Assaad}\,\orcidlink{0000-0002-3302-9243}}
\affiliation{Institut f\"ur Theoretische Physik und Astrophysik and W\"urzburg-Dresden Cluster of Excellence ctd.qmat, Universit\"at W\"urzburg, 97074 W\"urzburg, Germany}

\begin{abstract}
    At criticality, a local field acting at a single point in Euclidean space-time is irrelevant and induces a modulation in space and time that decays with twice its scaling dimension.
    This turns two-point correlation functions into one-point measurements: one operator insertion enters the Boltzmann weight, the other is measured. 
    For the local magnetization the relation is exact at any source strength.
    Implemented in auxiliary-field quantum Monte Carlo, the approach yields the static spin susceptibility and imaginary-time-displaced correlations with unprecedented precision.
    We apply it to the Dirac-semimetal to antiferromagnetic-Mott-insulator transition of the half-filled $\pi$-flux Hubbard model. 
    This lattice model provides a regularization of the $N_f=2$ Gross--Neveu--Heisenberg universality class.
    A finite-size scaling analysis estimates the critical exponents as
    $  \nu=1.10(3)$ and $\eta_\phi=1.09(4)$, close to field-theory predictions of $\eta_\phi\simeq1$.
    Since $\eta_\phi>1$, the equal-time structure factor at criticality does not diverge with the system size.
    Hence, the  equal-time structure factor at the ordering wave vector is dominated by background contributions and cannot be used for extracting critical exponents.
    Consistently, the spin dynamics at the transition is strongly incoherent and the quantum Fisher information remains bounded.
\end{abstract}

\maketitle
Gross--Neveu (GN) transitions describe the spontaneous generation of a mass for Dirac fermions through the breaking of a symmetry. 
Since the fermions remain gapless at the critical point, they qualitatively alter the critical behavior in contrast to purely bosonic order-parameter theories~\cite{Herbut09a}. 
Beyond graphene-type monolayers hosting Dirac fermions, such transitions may now be realized experimentally in WSe$_2$ tetralayers~\cite{Ma25}.
In renormalization-group theory~\cite{Cardy-book}, a
 critical point is characterized by the scaling dimensions
of its relevant operators, and considerable analytical and numerical effort has gone
into determining them. 
For the GN--Ising universality class, where an Ising symmetry is broken,
the different approaches agree well~\cite{Huffman19,Ihrig18,Gracey92,Knorr16,Erramilli23,wang2026}.
In contrast, for
the $N_f=2$ GN--Heisenberg (GNH)
universality class, a persistent discrepancy remains in the bosonic anomalous
dimension: perturbative and functional-renormalization-group calculations find
$\eta_\phi\simeq 1$~\cite{zerf2017,ladovrechis2023,gracey2018}, whereas fermionic
quantum Monte Carlo (QMC) studies based on equal-time correlations report
appreciably smaller values~\cite{parisen2015,otsuka2016,wang2026}.

In this Letter, we resolve this discrepancy by probing the critical point with a
source that couples to the order parameter at a single point in Euclidean space-time. 
Conventional pinning fields act along a line in imaginary time and thereby define a problem in the realm of boundary criticality~\cite{Assaad13,toldin2017}.
The point defect, in contrast, is irrelevant and does not alter bulk universality.
Within auxiliary-field QMC~\cite{Blankenbecler81,White89,Assaad08_rev}, 
it converts the evaluation of a two-point correlation function into that of the expectation value of a local operator,
and gives direct access to the static susceptibility.
Our estimate of the bosonic anomalous dimension agrees with field theory, and its value, $\eta_\phi>1$, explains
why previous QMC studies~\cite{parisen2015,otsuka2016,wang2026} based on equal-time correlations systematically underestimate it.

\begin{figure}[!t]
 \centering
 \includegraphics{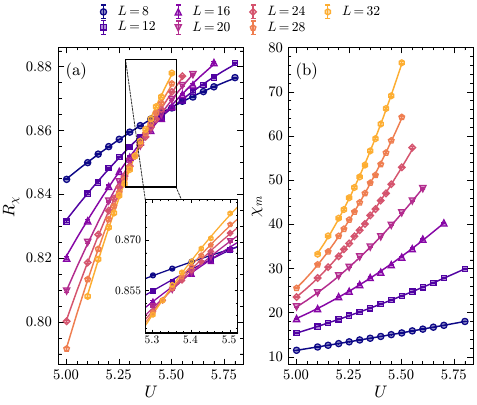}
 \caption{Finite-size scaling analysis of the $N_f=2$ GNH transition. (a) RG-invariant susceptibility ratio $R_\chi$ and (b) staggered spin susceptibility $\chi_m$ as a function of $U/t$ for $L=8$--$32$. Lines are polynomial fits to the data. }\label{fig:critical-exponents}
\end{figure}

Consider an action $S(\phi)$ supplemented by a local source term in space and time:
\begin{equation}
    S_h(\phi) = S(\phi) - h \int d^d \bm{x} \, d\tau \, \delta(\bm{x})\delta(\tau)   \phi(\bm{x},\tau).
\end{equation}
At a fixed point, where $S(\phi)$ is scale invariant, under a renormalization-group transformation the coupling transforms as
$h\to b^{-\Delta_\phi}h$, where $\Delta_\phi$ is the scaling dimension of the field $\phi$, and $b > 1$ is the scale factor of the transformation.
Since unitarity requires $\Delta_\phi>0$, the point defect is irrelevant.
This type of defect differs substantially from earlier pinning-field approaches defined at the Hamiltonian level.
The latter corresponds to an extended line defect in Euclidean space-time and has the potential of generating novel boundary critical exponents~\cite{toldin2017}.

In the presence of the space-time point defect, the expectation value of the field $\phi$ at position $\bm{x}$ and imaginary time $\tau$ is given by
\begin{equation}
 \begin{aligned}
 \langle\phi(\bm{x},\tau)\rangle_h
 &=\frac{\langle\phi(\bm{x},\tau)e^{h\phi(\bm{0},0)}\rangle_0}
 {\langle e^{h\phi(\bm{0},0)}\rangle_0}\\
 &= h\,\langle\phi(\bm{x},\tau)\phi(\bm{0},0)\rangle_0 + \mathcal{O}(h^3),
 \end{aligned}
 \label{eq:profile}
\end{equation}
where we assume $\langle\phi\rangle_0=0$. 
Since the defect is irrelevant, the effective coupling vanishes at long distances, 
and the profile remains proportional to the two-point function even at finite $h$, up to a renormalized amplitude. 
For the local magnetization this proportionality is in fact exact at any $h$, see Eq.~(\ref{eq:exact}) below. 
The defect thus turns a two-point function into a one-point measurement, which is advantageous in fermion QMC for two reasons:
First, the full space-time profile follows from equal-time Green's functions on each time slice, so
that no time-displaced Green's functions are required.
Second, two-point estimators involve products of Green's functions and therefore suffer from fat-tailed distributions near configurations where the fermion determinant vanishes~\cite{shi2016,ulybyshev2022,Liu21}.

\begin{figure}[t]
 \centering
 \includegraphics{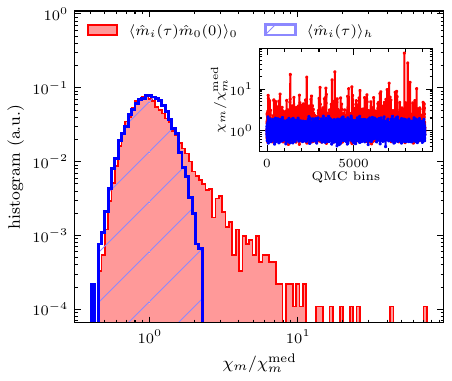}
\caption{Staggered spin susceptibility $\chi_m$, Eq.~(\ref{eq:susceptibility}), 
at $\bm{Q}=(\pi,\pi)$, $U=5.25\simeq U_c$, $L=\beta=24$, computed from the one-point correlator
 $\langle\hat m_{\bm{i}}(\tau)\rangle_h$ (blue, $h=1$) and from the two-point correlator 
 $\langle\hat m_{\bm{i}}(\tau)\hat m_{\bm{0}}(0)\rangle_0$ (red), which are related by Eq.~(\ref{eq:exact}). 
 Each QMC bin is the mean over 20 sweeps. 
 Both data sets are divided by their respective median $\chi_m^{\mathrm{med}}$ so as to remove the prefactor $C(U,L)$ in Eq.~(\ref{eq:exact}). 
 Main panel: histogram of the QMC bins, with the same number of bins for both estimators.
 The two-point estimator has rare events far above the median; the one-point estimator does not. 
 Inset: the same bins, $\chi_m/\chi_m^{\mathrm{med}}$ versus bin number. The two-point estimator shows spikes far from the median.}
 \label{fig:fat-tails}
\end{figure}

We apply this idea to the Hubbard model on the $\pi$-flux square lattice,
\begin{equation}
 \begin{aligned}
 \hat H={}&-t\sum_{\langle \bm{i}\bm{j}\rangle,\sigma}
 \left(e^{iA_{\bm{i}\bm{j}}}\hat c^\dagger_{\bm{i}\sigma}\hat c_{\bm{j}\sigma}+\mathrm{H.c.}\right)\\
 &+U\sum_{\bm{i}}\left(\hat{n}_{\bm{i}\uparrow}-\tfrac12\right)
                    \left(\hat{n}_{\bm{i}\downarrow}-\tfrac12\right).
 \end{aligned}
 \label{eq:model}
\end{equation}
Here, $\hat c^\dagger_{\bm{i}\sigma}$  creates a fermion with $z$ component of spin $\sigma$ in a Wannier state
centered around lattice site $\bm{i}$, $\hat n_{\bm{i}\sigma}=\hat c^\dagger_{\bm{i}\sigma}\hat c_{\bm{i}\sigma}$, with $\sigma=\uparrow,\downarrow$, 
and the phases satisfy $\sum_{\square}A_{\bm{i}\bm{j}}=\pi$. The weak-coupling phase is a
Dirac semimetal, and beyond $U_c/t$ the system becomes an antiferromagnetic
Mott insulator~\cite{Assaad13,parisen2015,otsuka2016}. 
The continuous transition belongs to the GNH universality class with
$N_f=2$ flavors of four-component Dirac fermions~\cite{Herbut09a,parisen2015,otsuka2016}.
At half filling, the model is free of the negative sign problem~\cite{Wu04}. This
remains true in the presence of the defect: under the partial particle-hole transformation
$\hat c_{\bm{i}\downarrow}\to(-1)^{\bm{i}}\hat c^\dagger_{\bm{i}\downarrow}$, the source
$h\hat m_{\bm 0}(0)$ maps onto $h(\hat n_{\bm 0}(0)-1)$, which couples
identically to both spin species, so the Boltzmann weight remains a product of
complex-conjugate determinants.

\noindent
\textit{Numerical method.}---We use the ALF implementation~\cite{assaadALF} of the
finite-temperature auxiliary-field QMC algorithm on $L\times L$ lattices with
periodic boundary conditions and $L=8,12,\dots,32$, and set $t=1$. We employ a
symmetric Trotter decomposition with $\Delta\tau =0.1$, a Hubbard--Stratonovich
decoupling in the density channel, and $\beta=L$, consistent with a dynamical critical exponent of $z=1$. Since the
algorithm is based on a Euclidean path integral, the point defect is implemented by
inserting the operator $e^{h\hat m_{\bm 0}}$, with
$\hat m_{\bm{i}}=\hat n_{\bm{i}\uparrow}-\hat n_{\bm{i}\downarrow}$, at the time slice $\tau=0$. 
The coupling $h$ is thus dimensionless and independent of $\Delta\tau$; we use $h=1$.

\noindent
\textit{Numerical results.}---
Since $\hat m_{\bm{i}}^3=\hat m_{\bm{i}}$, and since expectation values odd in $\hat m$, such as
$\langle\hat m_{\bm{0}}^{2n+1}\rangle_0$ and
$\langle\hat m_{\bm{i}}(\tau)\hat m_{\bm 0}^{2}(0)\rangle_0$, vanish due to time-reversal
symmetry, the relation between the one-point function and the zero-field two-point
function can be determined exactly:
\begin{equation}
 \langle \hat{m}_{\bm{i}}(\tau)\rangle_h= C(U,L)
 \langle \hat{m}_{\bm{i}}(\tau)\hat{m}_{\bm{0}}(0)\rangle_0,
 \label{eq:exact}
\end{equation}
with $C(U,L) = \frac{\sinh h}{1+(\cosh h-1)\langle \hat{m}_{\bm{0}}^2\rangle_0}$.
$C(U,L)$ is independent of $\bm{i}$ and $\tau$ and therefore does not affect the
profile of the correlation function, but it depends on $U$ and $L$ through
$\langle\hat m_{\bm 0}^2\rangle_0=\tfrac12-2\langle(\hat n_{\bm 0\uparrow}-\tfrac12)
(\hat n_{\bm 0\downarrow}-\tfrac12)\rangle_0$.

From the defect-induced profile we compute
\begin{equation}
 \chi(\bm q,0)=\int_0^\beta d\tau\sum_{\bm{i}} e^{i\bm q\cdot\bm i}
                         \langle \hat m_{\bm{i}}(\tau)\rangle_h ,
 \label{eq:susceptibility}
\end{equation}
which, by Eq.~(\ref{eq:exact}), is proportional to the static spin susceptibility at
wave vector $\bm q$, but is obtained from one-point measurements only. 
Within auxiliary-field QMC, Fig.~\ref{fig:fat-tails} illustrates the superiority of
the one-point estimator over the conventional two-point estimator: The one-point estimator does not
suffer from fat-tailed distributions.

At a critical point, $C(U,L)$ scales to a finite value and contributes to corrections to scaling.
The singular part of $\langle \hat{m}_{\bm{0}}^2\rangle_0$ in the vicinity of
criticality scales as $\partial f_s/\partial U \sim L^{1/\nu - d - z}$, where $f_s$ is
the singular part of the free energy density, $\nu$ the correlation length
exponent, $d=2$ the spatial dimension and $z=1$ for Lorentz-invariant fixed points. Since $d+z-1/\nu=\Delta_\epsilon >0$
is the scaling dimension of the energy operator, the denominator will contribute to corrections to scaling.
In our case, $\nu\approx 1$, such that $\Delta_\epsilon \approx 2$ and the corrections to scaling are
expected to be weak.

To locate
the transition, we use the RG-invariant ratio
\begin{equation}
 \begin{aligned}
 R_\chi&=1-
 \frac{\chi(\bm Q+\Delta\bm q,0)}{\chi(\bm Q,0)},\\
 \bm Q&=(\pi,\pi),\qquad |\Delta\bm q|=2\pi/L.
 \end{aligned}
 \label{eq:ratio}
\end{equation}
With $\beta=L$, finite-size scaling gives
\begin{equation}
 \begin{aligned}
 \chi_m(U,L)&=L^{2-\eta_\phi}\left[f_\chi(x)+L^{-\omega}g_\chi(x)\right],\\
 R_\chi(U,L)&=f_R(x)+L^{-\omega}g_R(x),
 \end{aligned}
 \label{eq:fss}
\end{equation}
with $x=(U-U_c)L^{1/\nu}$ and $\omega$ the leading correction-to-scaling exponent. To expose the $U,L$ dependence of the staggered spin susceptibility, we define $\chi_m(U,L) = \chi(\bm Q,0)$. Both quantities of Eq.~(\ref{eq:fss}) are plotted in Fig.~\ref{fig:critical-exponents}.

We extract the critical parameters
by a finite-size scaling analysis,
following Ref.~\cite{parisen2015}.
In the vicinity of the critical point, we can expand the right-hand side of Eq.~(\ref{eq:fss}) in powers of the scaling variable $x$:
\begin{gather}
	\begin{split}
 		R_\chi(U,L)=R^{*}&+\sum_{n=1}^{n_{\max}}a_n\,(U-U_c)^nL^{n/\nu}\\
 		                           &+L^{-\omega}\sum_{m=0}^{m_{\max}}\tilde{a}_m\,(U-U_c)^mL^{m/\nu},
 		\label{eq:fitR}
 	\end{split}\\
 \chi_m(U,L)=L^{2-\eta_\phi}\sum_{n=0}^{n_{\max}}b_n\,(U-U_c)^nL^{n/\nu}.
 \label{eq:fitchi}
\end{gather}
Neglecting for the moment corrections to scaling, we fit the QMC data to Eqs.~(\ref{eq:fitR}) and (\ref{eq:fitchi}) setting $\tilde{a}_m=0$, and using data points satisfying $|x|\le 4$.
To monitor the relevance of scaling corrections, fits are repeated for $L\ge L_{\min}$.
Fit results are reported in Table~\ref{tab:fits} in the End Matter.
For both observables, increasing $n_{\max}$ from $n_{\max}=2$ to $n_{\max}=3$ gives a minimal decrease of the $\chi^2/\operatorname{DOF}$ ($\operatorname{DOF}$ denotes the degrees of freedom), 
indicating that a quadratic approximation adequately captures the scaling functions $f_\chi$ and $f_R$.
As anticipated in Eq.~(\ref{eq:fitchi}), scaling corrections in $\chi_m$ appear to be negligible:
The fits show a good $\chi^2/\operatorname{DOF}$ and the fitted parameters are stable on increasing $L_{\min}$.
Fits of $R_\chi$ instead show a large $\chi^2/\operatorname{DOF}$, which persists on increasing $L_{\min}$, indicating that scaling corrections are numerically relevant.
To take them into account, we repeated the analysis including the correction term $\propto L^{-\omega}$ in Eq.~(\ref{eq:fitR}), leaving $\omega$ as a free parameter.
In order to better constrain the critical parameters, we performed a joint fit of $R_\chi$ and $\chi_m$ to Eq.~(\ref{eq:fitR}) and Eq.~(\ref{eq:fitchi}).
Fits for $n_{\max}=2$ and $m_{\max}=1$ display a good $\chi^2/\operatorname{DOF}$ and the fitted parameters are stable on increasing the minimum lattice size.
Furthermore, fitted values of $U_c$ and $\eta_\phi$ agree with those obtained from individual fits of $\chi_m$.
Inspecting the fit results, we arrive at the conservative estimates
\begin{equation}
 \begin{gathered}
 U_c=5.26(3),\\
 \nu=1.10(3),\qquad \eta_\phi=1.09(4).
 \end{gathered}
 \label{eq:final_results}
\end{equation}
These values and uncertainties are chosen to be in full agreement with joint fits and fits of $\chi_m$ in Table~\ref{tab:fits} with $\chi^2/\operatorname{DOF} \leq 1.5$ and $\omega \leq 2$. The latter criterion is an upper bound due to the presence of scaling corrections $\propto L^{-(2-\eta_\phi)}$.
The corresponding data collapse and an
independent crossing analysis are presented in Figs.~\ref{fig:data-collapse} and \ref{fig:crossing} in the End Matter.

\begin{figure}[t]
 \centering
 \includegraphics[width=0.45\textwidth]{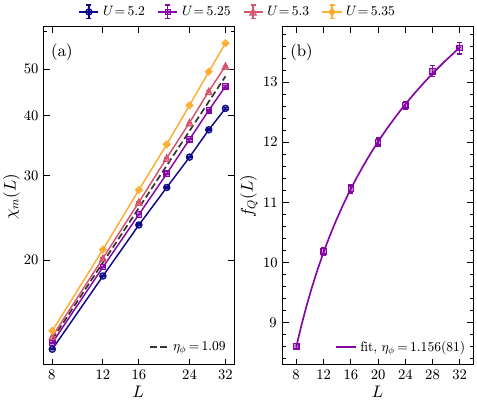}
 \caption{(a) Staggered spin susceptibility $\chi_m$ as a function of $L$ at $U=5.2$,
 $5.25$, $5.3$, and $5.35$ (log--log scale). The dashed line corresponds to
 $\chi_m\propto L^{2-\eta_\phi}$ with $\eta_\phi=1.09$. (b) Spin quantum Fisher
 information $f_Q$ as a function of $L$ at $U=5.25\simeq U_c$. 
 The uncertainties are determined via bootstrap resampling of the stochastic analytic continuation. The data was fitted to $A + BL^{1 - \eta_\phi}$ yielding $\eta_\phi = 1.156(81)$, consistent with the estimate in Eq.~\eqref{eq:final_results}. 
 }
 \label{fig:checks}
\end{figure}

Figure~\ref{fig:checks}(a) shows $\chi_m$ as a function of $L$ on a log--log scale
for four couplings around the critical point. For $U=5.2$ and $5.35$ the curves
bend downward and upward, respectively, bracketing $U_c$.
At $U=5.25$, close to $U_c$, the data follow a power law, consistent with $\eta_\phi=1.09$ (cf.~Eq.~(\ref{eq:final_results})).

This result also explains the smaller values of $\eta_\phi$ found in previous equal-time
analyses. At criticality, the singular part of the equal-time structure factor
scales as $S(\bm Q)\sim L^{1-\eta_\phi}$ and thus decreases with system size
for $\eta_\phi>1$, so that $S(\bm Q)$ is dominated by its non-universal
background. 
Hence, older results based on the equal-time structure factor have a larger estimate of $U_c$ than those presented in this work~\cite{parisen2015,otsuka2016}.

\begin{figure}[t]
 \centering
 \includegraphics{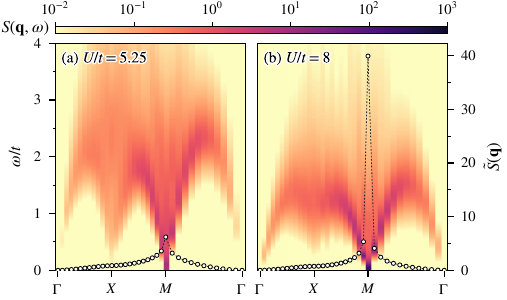}
 \caption{Dynamical spin structure factor $S(\bm q,\omega)$ along the path
 $\Gamma$--$X$--$M$--$\Gamma$, with $M=\bm Q=(\pi,\pi)$, for $L=\beta =24$ at
 (a) criticality, $U=5.25$, and (b) deep in the antiferromagnetic phase, $U=8$.
 The color scale is logarithmic. 
 The white dotted line (right axis, shared by both panels) shows 
 $\tilde{S}(\bm q) = S(\bm q) / \langle m_{\bm{0}} \rangle_h$, where 
 $S (\bm q)= \int d\omega\,S(\bm q,\omega)$ is the equal-time structure factor and 
 $\langle m_{\bm{0}} \rangle_h=0.6257(2)$ for $U=5.25$ and $0.6959(3)$ for $U=8$.
 Spectra are obtained by stochastic analytic continuation using the full covariance matrix
 of the imaginary-time data. }
 \label{Fig:spin-dynamics}
\end{figure}

\noindent
\textit{Spin dynamics and quantum Fisher information.}---Beyond the static
susceptibility, Eq.~(\ref{eq:exact}) yields the full imaginary-time dependence of
the spin correlations, $\langle\hat m_{\bm{i}}(\tau)\hat m_{\bm 0}(0)\rangle_0$, from
one-point measurements. We continue these data to real frequencies using the
stochastic analytic continuation~\cite{Sandvik98,Beach04a,Shao23} as implemented in
ALF~\cite{assaadALF}, taking into account the full covariance matrix of the
imaginary-time data.
 Figure~\ref{Fig:spin-dynamics} shows the dynamical spin structure factor
$S(\bm q,\omega)=\chi''(\bm q,\omega)/[\pi(1-e^{-\beta\omega})]$
at $U=5.25\simeq U_c$ and deep in the antiferromagnetic phase, $U=8$. 
At criticality, the spectral weight at $\bm Q$ 
exhibits a broad continuum at low frequencies, 
$S(\bm Q,\omega) \propto  \omega^{-(2-\eta_\phi)}$. 
In contrast, at $U=8$, spin-wave theory is valid and thus we expect a magnon zero-frequency pole at the ordering wave vector $\bm Q$.

The incoherent critical dynamics has a direct consequence for entanglement. The
quantum Fisher information density of the staggered magnetization,
\begin{equation}
 f_Q=\frac{4}{\pi N}\int_0^\infty d\omega\,\tanh\!\left(\frac{\beta\omega}{2}\right)
 \chi''(\bm Q,\omega),
 \label{eq:QFI}
\end{equation}
witnesses multipartite entanglement: since the local operator $\hat m_{\bm{i}}$ has
eigenvalues $0,\pm1$, $f_Q>4 C(U,L) k$ implies at least $(k+1)$-partite
entanglement~\cite{Hauke16}. The QFI is bounded by four times the variance,
$f_Q\le4S(\bm Q)$, with equality at $T=0$. For $\eta_\phi>1$ the singular part
of $S(\bm Q)$ decreases with system size, so $f_Q$ remains bounded at the
transition, as confirmed in Fig.~\ref{fig:checks}(b). A bounded QFI was also found at
Kondo-breakdown transitions~\cite{Mazza24}; in both cases, the critical spin dynamics
is incoherent and equal-time correlations do not diverge~\footnote{This conclusion relies on $\eta_\phi>1$. Our data support this, but the present error bars do not exclude  the possibility that $\eta_\phi$ falls below unity in larger-scale  simulations.}.
    
\noindent
\textit{Summary and conclusions.}---
We have shown that a source coupled to the order parameter at a single point in Euclidean space-time is an irrelevant
perturbation at criticality. 
Unlike extended defects, such as the imaginary-time
line defect generated by a Hamiltonian pinning field~\cite{toldin2017}, it does not
introduce an independent defect exponent, and the induced profile of the one-point measurement is
governed by the bulk two-point correlation function: one operator insertion enters the
Boltzmann weight, while the other is measured. 
For the magnetization, the relation
is exact at any $h$ (cf.~Eq.~(\ref{eq:exact})), so that spin correlations are accessible
also away from criticality; for other operators, it has to be checked on a case-by-case
basis. 
In auxiliary-field QMC, this approach provides a practical solution to 
the heavy-tailed distributions originating from the zeros of the fermion determinant~\cite{shi2016,ulybyshev2022}, 
such that large-scale simulations become well-behaved, due to the observed finite variance.
More broadly, point defects provide an
efficient route to space-time correlations and static susceptibilities, e.g., to
test emergent Lorentz symmetry, and supply imaginary-time data for analytic
continuation, as illustrated here for the critical spin dynamics. In particular, the
resulting dynamical susceptibility yields the quantum Fisher information, and thereby
provides a route to compute entanglement witnesses at quantum critical points with
fermion QMC. 
For the GNH transition, $\eta_\phi>1$ implies that the entanglement depth
witnessed by the QFI of the staggered magnetization remains finite at criticality. 

We applied the method to the GNH transition of the half-filled $\pi$-flux Hubbard
model and obtained $\nu=1.10(3)$ and $\eta_\phi=1.09(4)$, close to field-theory
predictions~\cite{zerf2017,ladovrechis2023,gracey2018}. Since $\eta_\phi>1$, the
singular part of equal-time spin correlations decreases with system size, so that
analyses based on them are dominated by non-universal background, whereas the
time-integrated susceptibility retains a growing singular part. This highlights the
difficulty of extracting critical behavior from equal-time correlations alone, when
the anomalous dimension is large. Finally, the method applies equally to the
GN--Ising~\cite{Huffman19,Wei16} and GN--XY~\cite{Otsuka18} universality classes;
as discussed in the End Matter, the point defect does not introduce a negative sign
problem in either case.

\begin{acknowledgments}
We thank Igor Herbut, João Inácio, Lukas Janssen, Fo-Hong Wang, Xiao Yan Xu, and Indra Yudhistira for useful discussions. We gratefully
acknowledge the Gauss Centre for Supercomputing e.V. for funding this project by
providing computing time on the GCS Supercomputer SuperMUC-NG at the Leibniz
Supercomputing Centre (LRZ) (project number pn73xu), as well as the scientific
support and HPC resources provided by the Erlangen National High Performance
Computing Center (NHR@FAU) of the Friedrich-Alexander-Universit\"at
Erlangen-N\"urnberg (FAU) under the NHR project b133ae. NHR funding is provided by
federal and Bavarian state authorities. NHR@FAU hardware is partially funded by the
German Research Foundation (DFG) -- 440719683. F.F.A., J.V., and L.T. acknowledge
financial support by the Deutsche Forschungsgemeinschaft (DFG, German Research
Foundation) through the W\"urzburg-Dresden Cluster of Excellence \textit{ctd.qmat}
-- Complexity, Topology and Dynamics in Quantum Matter (EXC 2147, Project
No.~390858490). F.P.T. is funded by the Deutsche Forschungsgemeinschaft (DFG,
German Research Foundation), Project No.~586743423.

\noindent
\textit{Use of AI.}--- We used several Claude models (Anthropic, 5.x generation), accessed through a Claude Team for Science subscription, to review the QMC code for errors and cross-check the data analysis. The authors directed all of this work, reviewed and verified the AI-generated output, and take full responsibility for the content of the paper.
\end{acknowledgments}

\bibliographystyle{apsrev4-2}
\bibliography{space_time_point_defect}

\clearpage
\section*{End Matter}

\begin{table*}[!t]
\caption{Finite-size-scaling fits of $R_\chi$ and $\chi_m$ to Eqs.~\eqref{eq:fitR} and~\eqref{eq:fitchi}. Rows above the horizontal line are individual fits without corrections to scaling; rows below are joint fits of $R_\chi$ and $\chi_m$ that include corrections to scaling for $R_\chi$. All fits are restricted to data with $|x| = |U-U_c|L^{1/\nu} \leq 4$; $L_{\min}$ refers to the smallest lattice size taken into account in the fits.}\label{tab:fits}
\begin{ruledtabular}
\begin{tabular}{cccccccc}
 & $L_{\min}$ & $U_c$ & $\nu$ & $\eta_\phi$ & $\omega$ & $\operatorname{DOF}$ & $\chi^2/\operatorname{DOF}$ \\
\hline
\multirow{3}{*}{$\begin{array}{c} R_\chi \\ n_{\max}=2 \end{array}$}
 & 8  & 5.4354(13) & 1.2350(88) & -- & -- & 80 & 57.59 \\
 & 12 & 5.3803(15) & 1.194(14)  & -- & -- & 65 & 15.40 \\
 & 16 & 5.3496(19) & 1.171(18)  & -- & -- & 52 & 4.21 \\[1ex]
\multirow{3}{*}{$\begin{array}{c} R_\chi \\ n_{\max}=3 \end{array}$}
 & 8  & 5.4285(12) & 1.1748(84) & -- & -- & 79 & 55.06 \\
 & 12 & 5.3813(15) & 1.198(14)  & -- & -- & 64 & 15.41 \\
 & 16 & 5.3496(19) & 1.171(18)  & -- & -- & 51 & 4.29 \\[1ex]
\multirow{3}{*}{$\begin{array}{c} \chi_m \\ n_{\max}=2 \end{array}$}
 & 8  & 5.2644(28) & 1.0532(33) & 1.0954(27) & -- & 79 & 4.10 \\
 & 12 & 5.2535(61) & 1.0848(76) & 1.1097(73) & -- & 64 & 1.44 \\
 & 16 & 5.274(12)  & 1.112(14)  & 1.081(16)  & -- & 51 & 1.24 \\[1ex]
\multirow{3}{*}{$\begin{array}{c} \chi_m \\ n_{\max}=3 \end{array}$}
 & 8  & 5.2698(31) & 1.0473(35) & 1.0906(30) & -- & 78 & 3.56 \\
 & 12 & 5.2528(62) & 1.0855(75) & 1.1105(74) & -- & 63 & 1.45 \\
 & 16 & 5.270(12)  & 1.111(13)  & 1.085(16)  & -- & 50 & 1.14 \\
\hline
\multirow{3}{*}{$\begin{array}{c} R_\chi+\chi_m \\ n_{\max}=2,\ m_{\max}=0 \end{array}$}
 & 8  & 5.2939(20) & 1.0243(36) & 1.0685(20) & 0.758(29) & 159 & 8.88 \\
 & 12 & 5.2813(46) & 1.0619(80) & 1.0775(56) & 1.27(11)  & 129 & 2.21 \\
 & 16 & 5.2955(95) & 1.106(15)  & 1.051(13)  & 2.6(26)   & 103 & 1.09 \\[1ex]
\multirow{3}{*}{$\begin{array}{c} R_\chi+\chi_m \\ n_{\max}=2,\ m_{\max}=1 \end{array}$}
 & 8  & 5.2609(28) & 1.0601(35) & 1.0984(27) & 0.338(24) & 158 & 4.20 \\
 & 12 & 5.2541(62) & 1.0924(79) & 1.1086(73) & 0.725(86) & 128 & 1.37 \\
 & 16 & 5.279(15)  & 1.120(15)  & 1.073(20)  & 1.6(22)   & 102 & 1.03 \\
\end{tabular}
\end{ruledtabular}
\end{table*}

\section{GN--Ising universality class}
A canonical lattice model for the GN--Ising universality class is the spinless
$t$-$V$ model on the honeycomb or $\pi$-flux lattice,
\begin{equation}
\hat H = -t \sum_{\langle \bm{i},\bm{j}\rangle} \left(\hat{c}_{\bm{i}}^\dagger \hat{c}_{\bm{j}} + \text{H.c.}\right)
 + V \sum_{\langle \bm{i},\bm{j}\rangle} \left(\hat{n}_{\bm{i}} - \tfrac12\right) \left(\hat{n}_{\bm{j}} - \tfrac12\right).
\end{equation}
This model is free of the negative sign problem~\cite{Huffman14,Li16}, and an
auxiliary-field QMC implementation is described in Ref.~\cite{assaadALF}. For large
$V$, the ground state is a charge density wave with a density imbalance between the
two sublattices. As a point defect, we insert $h(\hat n_{\bm 0}-\tfrac12)$ at
$\tau=0$. Since $(\hat n_{\bm{i}}-\tfrac12)^2=\tfrac14$, one has
$e^{h(\hat n_{\bm 0}-1/2)}=\cosh(h/2)+2\sinh(h/2)(\hat n_{\bm 0}-\tfrac12)$,
and particle-hole symmetry, $\langle \hat n_{\bm{i}}-\tfrac12\rangle_0=0$, yields for any $h$
\begin{equation}
\langle \hat{n}_{\bm{i}}(\tau) - \tfrac12 \rangle_h = 2 \tanh\!\left( \frac{h}{2} \right)
 \left\langle \left(\hat{n}_{\bm{i}}(\tau) - \tfrac12\right)
 \left(\hat{n}_{\bm 0}(0) - \tfrac12\right)\right\rangle_0 .
\end{equation}
As for the Hubbard model, the defect thus gives exact access to the zero-field
two-point function. Majorana positivity~\cite{Wei16} guarantees that a staggered
field, and hence the point defect, does not generate a negative sign problem.

\begin{figure}[b]
 \centering
 \includegraphics{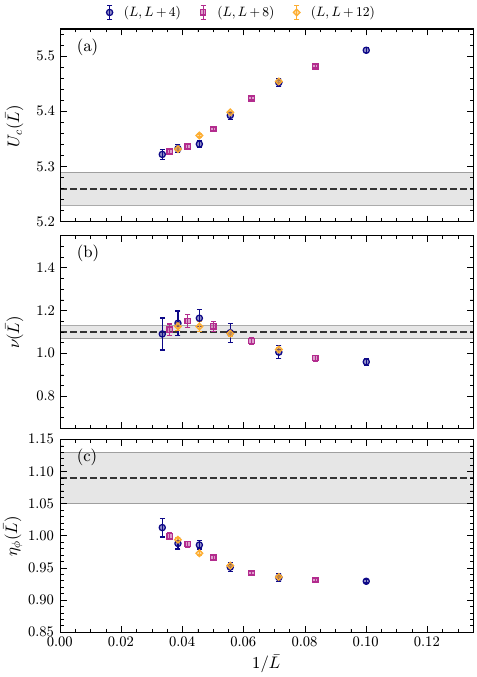}
 \caption{Crossing analysis: (a) $U_c$, (b) $\nu$, and (c) $\eta_\phi$ from
 Eq.~(\ref{eq:exponents}) for the pairs $(L,L+L_{\rm step})$ versus $1/\bar L$, with
 $\bar L$ the mean size of the pair. Dashed lines and shaded bands: estimates of
 Eq.~(\ref{eq:final_results}) with one-standard-deviation uncertainties.}\label{fig:crossing}
\end{figure}

\section{GN--XY universality class}
A realization of the GN--XY universality class was introduced in
Ref.~\cite{Otsuka18}: spinful fermions on the triangular lattice with an attractive
Hubbard interaction,
\begin{equation}
\hat H = \sum_{\langle \bm{i},\bm{j}\rangle, \sigma} t_{\bm{i}\bm{j}}\left(\hat{c}_{\bm{i}\sigma}^\dagger \hat{c}_{\bm{j}\sigma} + \text{H.c.}\right)
 - U \sum_{\bm{i}} \left(\hat{n}_{\bm{i}} -1 \right)^2 ,
\end{equation}
where $t_{\bm{i}\bm{j}}=\pm t$ accounts for a $\pi$-flux threading the upward-pointing
triangles and $U>0$. At half filling, corresponding to $\mu = 0$~\cite{Kennedy26},  the flux generates a Dirac semimetal. Since
the triangular lattice is not bipartite, the $SO(4)$ symmetry of the Hubbard model~\cite{Yang90}
on bipartite lattices is reduced to $U(1)\times SU(2)$, corresponding to charge
conservation and spin rotations. At a critical value of $U$, the model undergoes a
transition from the semimetal to an $s$-wave superconductor, which belongs to the
GN--XY universality class~\cite{Otsuka18}.

As a point defect, we insert the Hermitian local pairing operator
$h\hat O_{\bm 0}$, with
$\hat O_{\bm{i}}=\hat\Delta_{\bm{i}}+\hat\Delta_{\bm{i}}^\dagger$ and
$\hat\Delta_{\bm{i}}=\hat c_{\bm{i}\downarrow}\hat c_{\bm{i}\uparrow}$, at $\tau=0$. Since
$\hat O_{\bm{i}}^3=\hat O_{\bm{i}}$ and expectation values that change the particle number vanish
by $U(1)$ symmetry, the relation is exact and takes the same form as
Eq.~(\ref{eq:exact}):
\begin{equation}
 \langle \hat{O}_{\bm{i}}(\tau)\rangle_h=
 \frac{\sinh h}{1+(\cosh h-1)\langle \hat{O}_{\bm{0}}^2\rangle_0}
 \langle \hat{O}_{\bm{i}}(\tau)\hat{O}_{\bm{0}}(0)\rangle_0 .
\end{equation}

Interestingly, we were not able to demonstrate the absence of the negative sign
problem using the Majorana time-reversal symmetries of Ref.~\cite{Li16} and Majorana positivity~\cite{Wei16}.  Here we
follow a different route. Since $\hat O_{\bm 0}^2=(\hat n_{\bm 0}-1)^2$ is a projector,
$e^{h\hat O_{\bm 0}} = e^{ \alpha(\hat n_{\bm 0}-1)^2} + \sinh(h)\,\hat O_{\bm 0}$
with $\alpha = \ln(\cosh h) \ge 0$. After a discrete Hubbard--Stratonovich (HS)
decoupling of the perfect squares in the Hamiltonian, the Boltzmann weight of each
configuration $s$ reads
$\mathrm{Tr}\big[\hat{U}_{s}(\beta,0)\, e^{h \hat O_{\bm 0}}\big] =
\mathrm{Tr}\big[\hat{U}_{s}(\beta,0)\, e^{ \alpha(\hat n_{\bm 0}-1)^2}\big]$,
where $\hat{U}_{s}(\beta,0)$ is the imaginary-time propagator for the HS
configuration $s$. We have dropped the $\sinh(h)\,\hat O_{\bm 0}$ term since
$\hat{U}_{s}(\beta,0)$ conserves particle number. One can now carry out an
additional HS transformation of the space-time defect term. Since, for each
configuration of HS fields, the propagator is invariant under time reversal and
conserves particle number, Ref.~\cite{Wu04} guarantees the absence of the negative
sign problem. Summing over the additional field recovers
$\mathrm{Tr}\big[\hat U_s(\beta,0)\,e^{h\hat O_{\bm 0}}\big]$, which is therefore
non-negative for each $s$.

In practice, we carry out a partial particle-hole transformation on the down-spin
fermions,  $\hat c_{\bm{i}\downarrow}\to\hat c_{\bm{i}\downarrow}^\dagger$. It maps the
attractive interaction onto a repulsive one, reverses the sign of the down-spin
hopping, $t_{\bm{i}\bm{j}}\to-t_{\bm{i}\bm{j}}$, and maps $\hat O_{\bm{i}}$ onto the transverse magnetization
$\hat c_{\bm{i}\uparrow}^\dagger\hat c_{\bm{i}\downarrow}+\text{H.c.}$ The GN--XY case thus
reduces to a transverse-spin defect in a repulsive Hubbard model with
spin-dependent hopping, which can be treated exactly as in the main text.

\section{Crossing method}
At crossings of sizes $L$ and $rL$, the slopes of the ratio estimate $1/\nu$, while the susceptibility ratio estimates $\eta_\phi$. Writing $R_L=R_\chi(U,L)$ and $\chi_L=\chi_m(U,L)$, we use
\begin{equation}
\begin{aligned}
 \nu^{-1}(L)&=\left.\frac{\ln\!\left[\partial_U R_{rL}/\partial_U R_L\right]}{\ln r}\right|_{U=U_c(L)},\\
 \eta_\phi(L)&=2-\left.\frac{\ln\!\left[\chi_{rL}/\chi_L\right]}{\ln r}\right|_{U=U_c(L)}.
 \label{eq:exponents}
\end{aligned}
\end{equation}

\begin{figure}[H]
 \centering
 \includegraphics{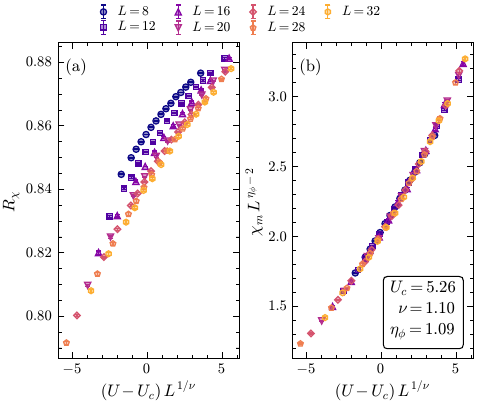}
 \caption{Data collapse of (a) $R_\chi$ and (b) $\chi_m$ using
 Eq.~(\ref{eq:final_results}): $U_c=5.26$, $\nu=1.10$, and $\eta_\phi=1.09$.}\label{fig:data-collapse}
\end{figure}

Figure~\ref{fig:crossing} shows the resulting finite-size estimates for the size
pairs $(L,L+L_{\rm step})$ with $L_{\rm step}=4$, $8$, and $12$; crossing
uncertainties are obtained by bootstrap resampling of Monte Carlo bins. As the
mean size $\bar L $ of the pair increases, the crossings drift toward smaller
couplings, reaching $U\simeq5.3$ for the largest pairs, while $\eta_\phi$ grows
and exceeds unity. Both trends point toward the estimates of Eq.~(\ref{eq:final_results}), shown as shaded bands. The large uncertainties in the finite-size estimates of $\nu$ arise because Eq.~(\ref{eq:exponents}) requires not only the interpolated value of $R_\chi$ at the crossing but also its derivative $\partial_U R_\chi$. 
Finally, Fig.~\ref{fig:data-collapse} shows the data collapse of the RG-invariant susceptibility ratio $R_\chi$ and the staggered susceptibility $\chi_m$ using the estimates of Eq.~(\ref{eq:final_results}). The susceptibility collapses for all sizes, while the ratio shows visible corrections to scaling for the smallest sizes.

\end{document}